\documentclass[aps,10pt,prd,superscriptaddress,preprintnumbers,%
showkeys,nofootinbib,showpacs,twocolumn]{revtex4-2}
\usepackage{hyperref}
\usepackage{graphicx}
\usepackage{amsmath}
\usepackage{bm}
\usepackage{slashed}
\usepackage{amssymb}
\begin{document}

\title{Complete definition of $N \rightarrow \Delta$ transition generalized parton distributions}

\author{June-Young Kim}
\affiliation{Theory Center, Jefferson Lab, Newport News, VA 23606, USA}
\author{Kirill~M.~Semenov-Tian-Shansky}
\affiliation{Department of Physics, Kyungpook National University, Daegu 41566, Korea}
\affiliation{NRC ``Kurchatov Institute'' - PNPI, Gatchina 188300, Russia}
\affiliation{Higher School of Economics, National Research University, 194100 St. Petersburg, Russia}
\author{Ho-Yeon Won}
\affiliation{CPHT, CNRS, \'Ecole Polytechnique, Institut Polytechnique de Paris, 91120 Palaiseau, France}
\author{Sangyeong Son}
\affiliation{Department of Physics, Kyungpook National University, Daegu 41566, Korea}
\author{Christian~Weiss}
\affiliation{Theory Center, Jefferson Lab, Newport News, VA 23606, USA}
\begin{abstract}
We revisit the definition of the leading-twist chiral-even generalized parton distributions (GPDs)
for $N \to \Delta$ baryon transitions. We identify and address deficiencies in
previous definitions of the transition GPDs inspired by the transition form factors of the
vector and axial-vector currents. Through systematic analysis of all possible covariant structures,
respecting discrete symmetries and the baryon spinor equations of motion, we derive complete sets of
independent structures for the transition matrix elements of the vector and axial-vector partonic operators.
They contain additional structures proportional to the light-cone vector, corresponding to
transition GPDs of vanishing first moment, which were not included in previous parametrizations.
Their presence is confirmed independently by the light-front multipole expansion and the cross-channel
SO(3) partial-wave analysis of the transition matrix elements. Our analysis provides a complete
definition of the $N \to \Delta$ transition GPDs for use in theoretical and phenomenological studies.
\end{abstract}
\maketitle
\tableofcontents
\section{Introduction}
Generalized parton distributions (GPDs) are a powerful tool for characterizing the structure of the
nucleon in QCD. They parametrize the nucleon matrix elements of partonic QCD
operators, describing the correlations of quark and gluon fields at light-like distances,
at nonzero momentum transfer between the nucleon states. 
They unify the concepts of the nucleon parton densities and elastic form factors
and provide essential new information on nucleon structure;
see Refs.~\cite{Ji:1996nm, Goeke:2001tz,Belitsky:2005qn, Diehl:2003ny, Boffi:2007yc} for a review. 
This includes the spatial distributions of quarks/gluons in the nucleon and the mechanical properties
such as the distributions of mass, angular momentum, and forces in system.
The GPDs are probed in lepton-nucleon scattering at energy/momentum transfers $\gg 1$ GeV with
exclusive final states, $l + N \rightarrow l' + M + N$ $(M = \textrm{photon, meson})$, where QCD factorization
methods can be used to describe the reaction in terms of quark/gluon processes.
The nucleon GPDs have been the object of extensive theoretical and experimental studies.

The concept of GPDs can be extended to transitions between the nucleon and other baryon states.
Transition GPDs describe the matrix elements of the partonic QCD operators for transitions
$N \rightarrow \pi N, \Delta, N^\ast$ etc., where the final state can be nonresonant or resonant.
They open up interesting new applications to hadronic physics. Transition GPDs enable the study
of baryon resonance excitation with new classes of QCD operators, greatly expanding the information
available from the transition form factors of the electromagnetic current operator; see
Refs.~\cite{Pascalutsa:2006up,Ramalho:2023hqd} for a review. They can also be used to characterize
the spatial distributions of quarks/gluons and the mechanical properties of baryon resonances.
The transition GPDs are probed in exclusive processes with baryon transitions
$l + N \rightarrow l' + M + N^\ast$ (or $\pi N, \Delta$), in the same kinematics as the
$N \rightarrow N$ exclusive processes. A program of transition GPD studies is emerging,
combining theoretical and experimental efforts \cite{Diehl:2024bmd}.

Of particular interest are the $N \rightarrow \Delta$ transition GPDs. Because of the isospin
difference between the $N$ and $\Delta$, the transition GPDs select the isovector components
of the quark partonic operators, enabling their separate study. $N\rightarrow \Delta$ transitions
in QCD can be described using systematic methods based on the $1/N_c$ expansion, providing reliable
predictions for the transition GPDs. As a strong resonance in the P-wave $\pi N$ channel, the
$\Delta$ is easy to reconstruct experimentally in the final state of hard exclusive processes.
Measurements of hard exclusive pion production with $N \rightarrow \Delta$ transitions are
being performed at JLab 12 GeV \cite{Diehl:2024bmd, CLAS:2023akb};
measurements of photon production (deeply-virtual Compton scattering) with
$N \rightarrow \Delta$ transitions are planned \cite{Semenov-Tian-Shansky:2023bsy}.

The parametrization of the $N \rightarrow \Delta$ transition matrix elements of the
partonic QCD operators and the definition of the transition GPDs pose some specific challenges.
The relativistic description of the spin-3/2 particle employs vector-bispinor wave functions
with constraints and requires a careful procedure for constructing the bilinear forms
describing the spin-1/2 to 3/2 transitions.
These questions have been addressed in the definition of the transition form factors
of the local vector and axial vector currents, but new issues arise in the case of
the GPDs that require separate treatment.
Several publications use incomplete definitions of the $N\rightarrow \Delta$ transition
GPDs inherited from the transition form factors \cite{Goeke:2001tz, Belitsky:2005qn}.

In this work we revisit the definition of the $N \rightarrow \Delta$ transition GPDs and
provide complete parametrizations of the transition matrix elements of the chiral-even
vector and axial vector operators. We review the construction of bilinear forms in the
parametrization of the vector and axial vector current matrix elements and point out the
differences arising in the case of partonic operators. We identify the issues with
existing parametrizations of the transition GPDs \cite{Goeke:2001tz, Belitsky:2005qn}
and propose a complete definition.

In Sec.~\ref{sec:transition_gpds} we define the $N\rightarrow \Delta$ matrix elements
of the partonic operators and describe the method for constructing parametrizations
by bilinear forms in the $N$ and $\Delta$ spinors. In Sec.~\ref{sec:vector} we examine
the definition of the vector transition GPDs. We review the parametrization of the
transition matrix elements of the electromagnetic current, discuss the issues with the
extension to the partonic operators and their treatment in the literature,
and propose a complete definition of the transition GPDs. In Sec.~\ref{sec:axial}
we perform the same analysis for the axial vector form factors and GPDs.
In Sec.~\ref{sec:interpretation} we interpret the new structures in the parametrization
from the perspective of the light-front multipole expansion, the $1/N_c$ expansion, and the
cross-channel partial wave expansion. In Sec.~\ref{sec:summary}
we present our conclusions and discuss possible extensions of the present study.
In Appendix~\ref{app:conventions} we explain the correspondence between the bilinear
forms obtained with different conventions of the antisymmetric tensor in the literature.

In the present study we focus on the parametrization of the $N \rightarrow \Delta$ transition matrix
elements and treat the $\Delta$ as a stable particle. Questions related to the definition of
resonance structure through complex analyticity, separation of resonant and non-resonant parts
of the $N \rightarrow \pi N$ transition matrix elements, finite $\Delta$ width, etc., remain
outside the scope and are discussed in Refs.~\cite{Diehl:2024bmd, Semenov-Tian-Shansky:2023bsy}.

\section{$N \rightarrow \Delta$ transition GPDs}
\label{sec:transition_gpds}
The transition matrix elements of the leading-twist chiral-even partonic QCD operators between
$N$ and $\Delta$ states are defined as
\begin{align}
&\mathcal{M}[\Gamma ]
=\int
\frac{d\tau}{2\pi} e^{i \tau x}
\langle \Delta  | \bar{\psi}(-\tau n/2) \Gamma^\mu n_\mu \hat{T} \psi(\tau n/2) | N \rangle .
\label{eq:mat_GPD}
\end{align}
The baryon states $|N\rangle \equiv |N(p_N, \lambda_N)\rangle$
and $|\Delta\rangle \equiv |\Delta(p_\Delta, \lambda_\Delta)\rangle$
are characterized by their 4-momenta $p_N$ and $p_\Delta$ and spin quantum numbers $\lambda_N$ and $\lambda_\Delta$.
The 4-momenta are on mass shell, $p_N^2 = m_N^2$ and $p_\Delta^2 = m_\Delta^2$, and the $\Delta$ is regarded
as a stable particle. The average baryon 4-momentum and the momentum transfer are defined as
\begin{align}
P \equiv \frac{p_{\Delta} + p_{N}}{2}, \hspace{2em} \Delta \equiv p_{\Delta} - p_{N}.
\label{P_Delta}
\end{align}
In the QCD operator in Eq.~(\ref{eq:mat_GPD}), $\psi$ and $\bar\psi$ are the quark fields,
and $n$ is a light-like 4-vector satisfying
\begin{align}
n^2 = 0, \hspace{2em}  n \cdot P = 1.
\label{normalization_n}
\end{align}
$\Gamma$ denotes one of the chiral-even bispinor matrices
\begin{subequations}
\label{Gamma}
\begin{alignat}{3}
\Gamma^\mu = && \gamma^\mu         & \hspace{2em} && \textrm{(vector)}, \\
             && \gamma^\mu\gamma_5 &              && \textrm{(axial vector)};
\end{alignat}
\end{subequations}
we refer to respective operators and matrix elements as vector and axial vector.
$\hat{T}$ is a quark flavor matrix in the space of two light flavors ($u, d$).
In the transition matrix element the $N$ state has isospin $1/2$, the $\Delta$ has isospin $3/2$,
and the operator is an isovector ($\Delta T = 1$). The isospin components of the matrix element
are related by isospin symmetry. In the following we consider the operators with
\begin{align}
\hat{T} = \tau^\pm, \tau^3; \hspace{2em} \tau^\pm \equiv \mp{\textstyle\frac{1}{\sqrt{2}}}(\tau^1 \pm i \tau^2),
\end{align}
where $\tau^a (a = 1,2,3)$ are the Pauli matrices. For these operators the 
isospin factors of the components are given 
by the vector addition (Clebsch-Gordan) coefficients,
$C_{\rm iso} \equiv \langle \textstyle{\frac{1}{2}} T_N 1 M| \textstyle{\frac{3}{2}} T_\Delta \rangle$, where $M = \pm 1, 0$ and $T_{N, \Delta}$ are the $N$ and $\Delta$ isospin projections.
Explicitly:
\begin{align}
\begin{array}{ll}
C_{\rm iso}   & \text{operator matrix element} 
\\[2ex]
1   
& \langle \Delta^{++} | [-\sqrt{2} \bar u d] | p\rangle,  \; \langle \Delta^{-} | \sqrt{2} \bar d u | n\rangle
\\[1ex]
\sqrt{\frac{2}{3}} \hspace{2em}
& \langle \Delta^{+} | \bar u u - \bar d d | p\rangle, \;
\langle \Delta^{0} |  \bar u u - \bar d d | n\rangle
\\[2ex]
\frac{1}{\sqrt{3}}
& \langle \Delta^0 | \sqrt{2} \bar d u | p\rangle, \;
\langle \Delta^+ | [-\sqrt{2} \bar u d] | n\rangle
\end{array}
\label{isospin_symmetry}
\end{align}
The convention for the overall sign and normalization of the matrix elements
agrees with the one of Ref.~\cite{Belitsky:2005qn}.
In the following we quote general expressions of the matrix elements valid for all isospin components;
in these expressions the factor $C_{\rm iso}$ appears explicitly and is to be taken at the value
of Eq.~(\ref{isospin_symmetry}) to get the matrix element of a given isospin component.

The matrix element Eq.~(\ref{eq:mat_GPD}) depends on the partonic variable $x$,
which is Fourier-conjugate to the light-like distance $\tau$.
It also depends on the invariant variables formed with the 4-momentum transfer $\Delta$,
\begin{align}
\xi \equiv - \frac{n\cdot \Delta}{2},
\hspace{2em}
t \equiv \Delta^2 .
\label{xi_t_def}
\end{align}
Finally it depends on the spin quantum numbers $\lambda_N$ and $\lambda_\Delta$. Altogether
\begin{align}
\mathcal{M}[\Gamma] 
= \textrm{function}(x, \xi, t; \lambda_\Delta, \lambda_N) .
\label{me_partonic_function}
\end{align}

The number of independent spin structures in Eq.~(\ref{me_partonic_function}) is constrained by symmetries
and can be established using general methods, such as the formalism of quark-hadron helicity
amplitudes \cite{Diehl:2003ny} or the transverse multipole expansion of the matrix element of the
partonic operator \cite{Kim:2024ibz}. Applying these methods to the $N\rightarrow\Delta$
transition matrix elements one determines the number of independent spin structures as
\begin{subequations}
\label{number_structures}
\begin{align}
\textrm{vector:} & & \textrm{4 structures},
\\[1ex]
\textrm{axial vector:} & & \textrm{4 structures}.
\end{align}
\end{subequations}
These numbers are confirmed by the explicit results of the covariant decomposition of the matrix elements
(see Secs.~\ref{sec:vector} and \ref{sec:axial}) and the multipole expansion and cross-channel SO(3)
partial wave analysis of matrix element (see Sec.~\ref{sec:interpretation}).

The dependence on the baryon spin variables can be made explicit by expanding the matrix element
Eq.~(\ref{eq:mat_GPD}) in bilinear forms in the $N$ and $\Delta$ spin wave functions.
The $N$ is described by a spin-1/2 bispinor (Dirac spinor) $u \equiv u(p_N, \lambda_N)$, satisfying
the dynamical equation
\begin{align}
(\slashed{p}{}_{N} - m_N) u &= 0.
\label{spinor_nucleon}
\end{align}
The $\Delta$ is described by a spin-3/2 vector-bispinor (Rarita-Schwinger spinor)
$u^\alpha \equiv u^{\alpha}(p_{\Delta},\lambda_{\Delta})$, satisfying a
similar dynamical equation and covariant constraints ensuring the projection
on spin 3/2 for on-shell momenta (absence of spin-1/2 components),
\begin{subequations}
\label{spinor_delta}
\begin{align}
&(\slashed{p}{}_{\Delta} - m_\Delta ) u^\alpha = 0,
\\
&\gamma^\alpha u_\alpha = 0, \hspace{2em} p_\Delta^\alpha u_\alpha = 0.
\end{align}
\end{subequations}
The matrix element Eq.~(\ref{eq:mat_GPD}) can then be represented as 
\begin{align}
\mathcal{M}[\Gamma]
= C_{\rm iso} \sum_{I = 1,2,3,4} \bar{u}_\alpha (\lambda_\Delta ) \mathcal{K}_I^{\alpha\mu} n_\mu u (\lambda_N)
\, H_{I}(x, \xi, t) .
\label{bilinear_partonic}
\end{align}
The functions $H_{I}(x, \xi, t)$ are the GPDs and represent the
invariant amplitudes associated with the matrix element Eq.~(\ref{eq:mat_GPD}).
The dependence on $\lambda_N$ and $\lambda_\Delta$ is contained in the bilinear forms
in the spinors. The spinor matrices
\begin{align}
\mathcal{K}_I^{\alpha\mu} n_\mu  \equiv \mathcal{K}_I^{\alpha\mu}(P, \Delta, n) \, n_\mu 
\end{align}
are constructed from the Dirac matrices and the independent 4-vectors $P, \Delta$ and $n$,
as well as from the invariant numerical tensors $g^{\mu\nu}$ and $\epsilon^{\mu\nu\rho\sigma}$,
and represent the various independent spin-spin and spin-orbit couplings allowed by the symmetries.
The definition of these structures is not unique and requires choices.
In the following we investigate the possible choices of these structures their
and consequences for the definition of the associated GPDs.

Integration over the partonic variable $x$ reduces the partonic operator in Eq.~(\ref{eq:mat_GPD})
to the local vector or axial vector current operator. Parallel to Eq.~(\ref{eq:mat_GPD}), we
therefore consider the transition matrix element of the local operator
\begin{align}
\mathcal{J}^\mu [\Gamma ]
&\equiv \langle \Delta  | \bar{\psi}(0) \Gamma^{\mu} \hat{T} \psi(0) | N \rangle ,
\label{local}
\end{align}
with $\Gamma$ defined as in Eq.~(\ref{Gamma}) and the isospin components defined as in
Eq.~(\ref{isospin_symmetry}). It is related to Eq.~(\ref{eq:mat_GPD}) by the ``sum rule''
\begin{align}
\int_{-1}^1  dx\, \mathcal{M} [\Gamma ]
= n_\mu \mathcal{J}^\mu [\Gamma ].
\end{align}
The matrix element Eq.~(\ref{local}) does not depend on the light-cone vector $n$.
The number of independent structures is therefore constrained by 3-dimensional rotational invariance.
In addition, the matrix element of the vector current ($\Gamma = \gamma^\mu$) is constrained
by current conservation; the axial vector current ($\Gamma = \gamma^\mu \gamma_5$) is
not conserved but induces a coupling to pions because of chiral symmetry breaking (PCAC);
these are dynamical features of the nonperturbative interactions in the hadronic states.
The expansion in bilinear forms now takes the form
\begin{align}
\mathcal{J}^\mu [\Gamma]
= C_{\rm iso} \sum_{I = 1,2,3}
\bar{u}_\alpha(\lambda_\Delta)  \mathcal{K}_I^{\alpha \mu} u(\lambda_N) \,
G_{I}(t), 
\label{local_bilinear}
\end{align}
where $G_I(t)$ are the invariant form factors (vector or axial vector), and the spinor matrices
are now constructed from the 4-vectors $P$ and $\Delta$ only,
\begin{align}
\mathcal{K}_I^{\alpha \mu} \equiv \mathcal{K}_I^{\alpha \mu}(P, \Delta).
\end{align}
The known bilinear expansions of the vector and axial vector transition currents can provide
guidance for the bilinear expansion of the matrix element of the
partonic operator Eq.~(\ref{bilinear_partonic}).
The construction must respect number of independent structures in the local and partonic operators
and the differences resulting from presence of light-cone vector in partonic matrix element
Eq.~(\ref{bilinear_partonic}). The choice of the bilinear structures determines the form of the
sum rules connecting the GPDs with the form factors of the local operators.

In the following we examine the parametrizations of the transition matrix elements or the
local currents and the partonic operators that have been proposed so far,
identify issues with the expansion of the partonic operator, and propose complete
expansions that have the correct number of independent structures and GPDs
and a simple connection of the GPDs with existing form factor definitions.

\section{Vector transition GPDs}
\label{sec:vector}
\subsection{JS form factors}
The $N\rightarrow \Delta$ transition matrix element of local vector current was
analyzed by Jones and Scadron (JS) \cite{Jones:1972ky}. These authors considered
the electromagnetic current for $p \rightarrow \Delta^+$ transitions.
In the case of two quark flavors ($u,d$) the electromagnetic current operator
is $J^\mu_{\rm em} = \bar\psi \gamma^\mu \hat{Q} \psi$
with quark charge matrix
\begin{align}
\hat{Q} = \frac{1}{6} \bm{1} + \frac{1}{2} \tau^{3},
\label{eq:charge}
\end{align}
and only the second (isovector) term contributes to the $N\rightarrow \Delta$ transition matrix element,
so that 
\begin{align}
\langle\Delta^+ | J^\mu_{\rm em}(0) | p \rangle
&= \langle\Delta^+ | \bar\psi (0) \gamma^\mu \frac{\tau^3}{2} \psi (0) |p \rangle
\nonumber \\
&= \frac{1}{2} \; \mathcal{J}^\mu [\gamma]_{p \rightarrow \Delta^+}
\label{eq:mat_EM}
\end{align}
in the convention of Eq.~(\ref{local}).

The bilinear form decomposition Eq.~(\ref{local_bilinear}) of the matrix element of the
local vector operator Eq.~\eqref{eq:mat_EM} is constrained by the discrete symmetries
(parity, time-reversal, hermiticity). Taking into account the relations resulting from
the conditions on the spinor wave functions Eqs.~(\ref{spinor_nucleon}) and
(\ref{spinor_delta}), four independent bispinor matrices are identified:
\begin{align}
\Delta^{\alpha} \Delta^{\mu} \gamma_{5}, \quad \Delta^{\alpha} \gamma^{\mu}\gamma_{5},
\quad  \Delta^{\alpha} P^{\mu} \gamma_{5}, \quad g^{\alpha\mu}  \gamma_{5}.
\label{eq:possible_EM}
\end{align}
The matrix $\gamma_{5} \equiv i \gamma^{0}\gamma^{1}\gamma^{2}\gamma^{3}$ appears because of
the parity difference and the spin difference of one between the $N$ and $\Delta$ baryon states.
[In diagonal $N \rightarrow N$ matrix elements, time-reversal symmetry and hermiticity typically
reduce the number of possible structures, but in the non-diagonal $N \to \Delta$ matrix elements
these constraints are not effective, allowing for the four structures in Eq.~\eqref{eq:possible_EM}.]
The condition of conservation of the electromagnetic current reduces the four
structures to three. The decomposition of the electromagnetic current matrix element Eq.~(\ref{eq:mat_EM})
is chosen as \cite{Jones:1972ky}
\begin{align}
& \langle\Delta^+ | J^\mu_{\rm em}(0) | p \rangle
\nonumber \\
&= \sqrt{\frac{2}{3}}
\sum_{I=1,2,3} \bar{u}_{\alpha}(\lambda_\Delta)   \mathcal{K}_{I}^{ \alpha \mu} u(\lambda_N) \, G_{I}(t) .
\label{JS_covariant_type}
\end{align}
The factor $\sqrt{2/3}$ is the isospin factor of the $p \rightarrow \Delta^+$ component of the
matrix element Eq.~(\ref{isospin_symmetry}) \cite{Jones:1972ky}. The bispinor matrices are
defined as 
\begin{subequations}
\label{eq:2}
\begin{align}
\mathcal{K}^{\alpha \mu}_{1}&=
\frac{1}{m_{N}} (\Delta^{\alpha} \gamma^{\mu} - \slashed{\Delta} g^{\alpha \mu})\gamma_{5},
\\
\mathcal{K}^{\alpha \mu}_{2}&=
\frac{2}{m^{2}_{N}} (\Delta^{\alpha} P^{\mu} - \Delta \cdot P g^{\alpha \mu})\gamma_{5},
\\
\mathcal{K}^{\alpha \mu}_{3}&=
\frac{1}{m^{2}_{N}}(\Delta^{\alpha} \Delta^{\mu} - \Delta^{2} g^{\alpha \mu})\gamma_{5},
\end{align}
\end{subequations}
and explicitly satisfy current conservation, as can be verified by contracting with $\Delta_{\mu}$.
The form factors $G_I(t)~(I = 1,2,3)$ are the dimensionless covariant-type form factors of the
$N\to\Delta$ transition, analogous to the Dirac and Pauli form factors of the $N\to N$ transition.

Alternatively, the matrix element Eq.~(\ref{eq:mat_EM}) can be parameterized in terms of multipole-type
form factors. Using the relations \cite{Jones:1972ky}
\begin{subequations}
\label{eq:Sign}
\begin{align}
&i \epsilon^{\alpha  \mu P \Delta} \doteq -m_{\Delta} m_{N} \mathcal{K}^{\alpha  \mu}_{1}
+ \frac{m^{2}_{N}}{2}\mathcal{K}^{\alpha  \mu}_{2}+\frac{m^{2}_{N}}{2} \mathcal{K}^{\alpha  \mu}_{3},
\label{eq:Sign_a} \\
&\epsilon^{\alpha \ P \Delta}_{\ \lambda}\epsilon^{\mu \lambda P \Delta} \gamma_{5}
\nonumber \\
&\doteq -\Delta \cdot p_{\Delta} \frac{m^{2}_{N}}{2} \mathcal{K}^{\alpha \mu}_{2}
+ P \cdot p_\Delta \, m^{2}_{N} \mathcal{K}^{\alpha \mu}_{3}, \label{eq:Sign_b} \\
&\Delta^{\alpha }(\Delta^{2} P^{\mu} - \Delta \cdot P \Delta^{\mu})
\gamma_{5}
\nonumber \\
&\doteq \Delta^{2} \frac{m^{2}_{N}}{2} \mathcal{K}^{\alpha \mu}_{2}
- \Delta \cdot P \, m^{2}_{N} \mathcal{K}^{\alpha \mu}_{3}, \label{eq:Sign_c}
\end{align}
\end{subequations}
where $\doteq$ denotes on-shell equality, the expansion Eq.~(\ref{JS_covariant_type}) can be
reorganized as
\begin{align}
& \langle\Delta^+ | J^\mu_{\rm em}(0) | p \rangle
\nonumber \\
&= \sqrt{\frac{2}{3}}
\sum_{I=M,E,C} \bar{u}_{\alpha}(\lambda_\Delta) \mathcal{K}_{I}^{\alpha\mu} u(\lambda_N) \, G^{*}_{I}(t).
\label{eq:Jones_ME}
\end{align}
The bispinor matrices are defined as
\begin{subequations}
\label{eq:1}
\begin{align}
  {\mathcal{K}}_{M}^{\alpha \mu} &=
   -i\frac{3(m_{\Delta}+m_{N})}{2m_{N}Q_{+}^{2}}
\varepsilon^{\alpha \mu P \Delta} ,
\\
{\mathcal{K}}_{E}^{\alpha \mu}
&= -\mathcal{K}_{M}^{\alpha \mu}
- \frac{6(m_{\Delta}+m_{N})}{m_{N} Z(t)}
\varepsilon^{\alpha \sigma P \Delta} \varepsilon_{\ \sigma}^{\mu \ P  \Delta} \gamma_5,
\\
{\mathcal{K}}_{C}^{\alpha \mu} &=-\frac{3(m_{\Delta}+m_{N})}{m_{N}Z(t)}
\Delta^{\alpha} (\Delta^2 P^{\mu} -\Delta\cdot P \Delta^{\mu}) \gamma_5,
\end{align}
\end{subequations}
with
\begin{align}
Z(t)&=Q^{2}_{-}Q^{2}_{+}, \hspace{2em} Q^{2}_{\pm} =[(m_{\Delta}\pm m_{N})^{2}-t].
\end{align}
Here the antisymmetric tensor is defined as\footnote{Some works in the literature
\cite{Pascalutsa:2006up,Semenov-Tian-Shansky:2023bsy,Goeke:2001tz} use a different
convention for the antisymmetric tensor. Here and in Sec.~\ref{sec:axial} we present
expressions in the convention of Eq.~(\ref{eq:conv_1}) even when referring to works
using the other convention. The correspondence between the bilinear forms in the
different conventions is explained in Appendix~\ref{app:conventions}.}
\begin{align}
&\epsilon^{0123}=-\epsilon_{0123}=1 ,
\label{eq:conv_1}
\end{align}
and the contraction with the momenta $P$ and $\Delta$ is denoted as
$\epsilon^{\alpha \mu P \Delta} \equiv \epsilon^{\alpha \mu \delta \lambda} P_{\delta} \Delta_{\lambda}$.
$G^{*}_{M,E,C}(t)$ are the magnetic $M1$, electric $E2$ and Coulomb $C2$ form factors.
The relationship between the two sets of electromagnetic form factors is given by \cite{Jones:1972ky}
\begin{widetext}
\begin{align}
&3m_{\Delta}m_{N} (m_{\Delta} + m_{N})
\left(\begin{array}{c} G^{*}_{M} \\ G^{*}_{E} \\  G^{*}_{C} \end{array}\right)
\nonumber \\
&= \left(\begin{array}{c c c}
[(3m_{\Delta}+m_{N})(m_{\Delta}+m_{N})-t] & 2(m_{\Delta}^{2}-m^{2}_{N}) & 2t
\\
( m^{2}_{\Delta}-m^{2}_{N}+t )  & 2(m_{\Delta}^{2}-m^{2}_{N})  & 2t
\\
4m^{2}_{\Delta} &  2(3m_{\Delta}^{2}+m^{2}_{N}-t)   & 2(m_{\Delta}^{2}-m^{2}_{N}+t)
\end{array} \right)
\left(\begin{array}{c} m_{N} G_{1} \\  m_{\Delta} G_{2} \\ m_{\Delta} G_{3} \end{array}\right).
\label{eq:connect}
\end{align}
\end{widetext}

In the above parametrizations the bilinear forms in the $N$ and $\Delta$ spinors are defined
on the mass shell, where the constraints Eq.~(\ref{spinor_delta}) project out the spin-3/2 part
of the vector-bispinor. In the context of effective field theory calculations involving $\Delta$
baryons, or in studies using the analytic properties of the matrix elements in $p_\Delta^2$, the
covariant structures in the parametrization must be defined off the mass-shell.
In the prescription of Ref.~\cite{Pascalutsa:2006up}, the use of so-called ``consistent''
couplings ensures the decoupling of the unphysical spin-1/2 degrees of freedom of the
Rarita-Schwinger field also off the mass shell. For the covariant structures this implies
the condition
\begin{align}
(p_{\Delta})_{\alpha} (\mathcal{K}^{\mathrm{con}}_I)^{\alpha \mu} (P,\Delta) = 0.
\label{eq:con}
\end{align}
A new decomposition of the vector current matrix element has been proposed, 
using structures which explicitly satisfy the
condition Eq.~(\ref{eq:con}) \cite{Semenov-Tian-Shansky:2023bsy}:
\begin{align}
&\langle\Delta^+ | J^\mu_{\rm em}(0) | p \rangle
\nonumber \\[1ex]
&= \sqrt{\frac{2}{3}} \sum_{I=M,E,C}
\bar{u}_{\alpha}(\lambda_\Delta) (\mathcal{K}^{\mathrm{con}}_{I})^{ \alpha \mu} u(\lambda_N)
\, g_{I}(t),
\label{eq:VdH_para1}
\end{align}
where
\begin{subequations}
\label{eq:VdH_para}
\begin{align}
(\mathcal{K}^{\mathrm{con}}_{M})^{\alpha \mu}
&= - \frac{3(m_{\Delta}+m_{N})}{2m_{N} Q^{2}_{+}} i\epsilon^{\alpha \mu p_{\Delta} \Delta},
\\[2ex]
(\mathcal{K}^{\mathrm{con}}_{E})^{\alpha \mu}
&= - \frac{3(m_{\Delta}+m_{N})}{2m_{N} Q^{2}_{+}}
\left( \Delta^{\alpha } p^{\mu}_{\Delta} - \Delta \cdot p_{\Delta} g^{\alpha \mu} \right) \gamma_{5},
\\[1ex]
(\mathcal{K}^{\mathrm{con}}_{C})^{\alpha \mu}
&= -\frac{3}{2} \frac{\left(m_\Delta+m_N\right)}{m_N m_\Delta Q_{+}^2}
\left[\gamma \cdot p_\Delta \left(\Delta^\alpha \Delta^\mu-\Delta^2 g^{\alpha \mu}\right)\right. 
\nonumber \\
& \hspace{1em} \left. -\gamma^\alpha\left(\Delta \cdot p_\Delta \,\Delta^\mu
-\Delta^2 p_\Delta ^\mu\right)\right] \gamma_5.
\end{align}
\end{subequations}
The new form factors are related to the covariant-type form factors as
\begin{align}
&3m_{\Delta}m_{N} (m_{\Delta} + m_{N})
 \left(\begin{array}{c}  g_{M} \\  g_{E}  \\   g_{C}  \end{array}\right) 
\nonumber \\
&=2 Q^{2}_{+} 
\left(\begin{array}{rrr} 1  & 0 & 0  \\
-1  & -2  & 0  \\
0 &  1   & -1    
\end{array} \right)  
\left(\begin{array}{c} m_{N} G_{1} \\  m_{\Delta} G_{2} \\ m_{\Delta} G_{3} \end{array}\right).
\label{eq:connect_2}
\end{align}
The relation to the multipole form factors is obtained by multiplying the conversion matrix
in Eq.~\eqref{eq:connect} with the inverse of the conversion matrix in Eq.~\eqref{eq:connect_2}
\cite{Pascalutsa:2006up}:
\begin{align}
&\left(\begin{array}{c} G^{*}_{M} \\ G^{*}_{E} \\  G^{*}_{C} \end{array}\right) 
= \left(\begin{array}{c c c}
1 & -\frac{m^{2}_{\Delta} - m^{2}_{N}+t}{2Q^{2}_{+}} & -\frac{t}{Q^{2}_{+}}  \\
0  & -\frac{m^{2}_{\Delta} - m^{2}_{N}+t}{2Q^{2}_{+}}  & -\frac{t}{Q^{2}_{+}}  \\
0 &  -\frac{2m^{2}_{\Delta}}{Q^{2}_{+}}  & -\frac{m^{2}_{\Delta} - m^{2}_{N}+t}{Q^{2}_{+}}
\end{array} \right)
\left(\begin{array}{c} g_{M} \\ g_{E} \\ g_{C} \end{array}\right).
\nonumber \\
\label{eq:connect_3}
\end{align}
\subsection{GPV parametrization}
A first parameterization of the $N\to \Delta$ transition matrix element of the partonic operator
Eq.~(\ref{eq:mat_GPD}) was introduced by Goeke, Polyakov and Vanderhaegehen (GPV) \cite{Goeke:2001tz}.
These authors built upon the heritage parametrization of the multipole-type electromagnetic form factors
Eq.~\eqref{JS_covariant_type} and represented the matrix element of the partonic operator as
\begin{align}
\mathcal{M}[\gamma]
= C_{\mathrm{iso}} \sum_{I=M,E,C} \bar{u}_{\alpha}(\lambda_\Delta)
\mathcal{K}_{I}^{\alpha\mu} n_{\mu} u(\lambda_N) \,
H_{I}(x,\xi,t) ,
\label{eq:Goeke_para}
\end{align}
where the tensors $\mathcal{K}^{\alpha\mu}_{M, E, C}$  are those of Eq.~\eqref{eq:1} and
$H_{M,E,C}$ are the multipole-type transition GPDs associated with these structures.
The first moments of these multipole GPDs are given the multipole form factors of Eq.~(\ref{eq:Jones_ME}),
\begin{align}
\int^{1}_{-1}  dx \, H_{M,E,C}(x,\xi,t) = 2G^{*}_{M,E,C}(t),
\label{eq:sum_1}
\end{align}
where the factor $2$ arises from the factor between the electromagnetic and the isovector current
in Eq.~\eqref{eq:mat_EM}.

The parametrization Eq.~\eqref{eq:Goeke_para} uses only three terms for the transition matrix element
of the partonic operator, which is one less than the number of independent spin structures determined
on general grounds, Eq.~(\ref{number_structures}). An additional structure is therefore required.
This also follows from the fact that the condition of current conservation, which reduced the
number of covariant structures in Eq.~(\ref{eq:possible_EM}) from four to three, does not apply
to the matrix element of the partonic operator. The structure in Eq.~\eqref{eq:possible_EM}
on grounds of current conservation should therefore be reinstated.

\subsection{BR parametrization}
An amended parametrization of the $N\to \Delta$ transition matrix element of the partonic
operator Eq.~(\ref{eq:mat_GPD}) was proposed by Belitsky and Radyushkin (BR) \cite{Belitsky:2005qn}.
It uses four bilinear structures, as required by the number of independent spin structures
Eq.~\eqref{number_structures},
\begin{align}
\mathcal{M}[\gamma] &= C_{\mathrm{iso}} \sum_{I=1,2,3,4}
\bar{u}_{\alpha}(\lambda_\Delta) \mathcal{K}_{I}^{\alpha\mu} n_{\mu} u(\lambda_N) \, G_{I}(x,\xi,t),
\label{eq:Belitsky_para}
\end{align}
where the tensors $\mathcal{K}^{ \alpha \mu}_{1, 2, 3}$ are those of Eq.~\eqref{eq:2},
and the additional fourth tensor is chosen as
\begin{align}
\mathcal{K}^{\alpha \mu}_{4}&= \frac{1}{m^{2}_{N}} \Delta^{\alpha} \Delta^{\mu} \gamma_{5};
\label{eq:fourth_term}
\end{align}
this term would violate current conservation in the matrix element of the local operator.
$G_{1,2,3,4}$ represent the covariant-type leading-twist vector GPDs. They are related to
the covariant-type form factors of Eq.~(\ref{JS_covariant_type}) by
\begin{subequations}
\label{eq:sum_2}
\begin{align}
\int^{1}_{-1}  dx \, G_{1,2,3}(x,\xi,t)&= 2G_{1,2,3}(t),
\\
\int^{1}_{-1}  dx \, G_{4}(x,\xi,t)&= 0.
\end{align}
\end{subequations}
The fourth GPD has zero integral because the corresponding structure is absent in the
matrix element of the local operator. The covariant-type GPDs $G_{1,2,3}$ of Eq.~(\ref{eq:Belitsky_para})
can be related to the mutipole-type GPDs of Eq.~(\ref{eq:Goeke_para}) by the same relations
as for the form factors in Eq.~\eqref{eq:connect}.

Note that Ref.~\cite{Belitsky:2005qn} uses a convention for the 4-momenta different from our
Eq.~(\ref{P_Delta}),
\begin{align}
P|_{\text{\cite{Belitsky:2005qn}}}= p_{N} + p_{\Delta} , \quad
\Delta|_{\text{\cite{Belitsky:2005qn}}} = p_{N} - p_{\Delta},
\label{eq:sign_dif}
\end{align}
which influences the definition of the covariant structures.
In Eqs.~(\ref{eq:Belitsky_para}) and Eq.~(\ref{eq:fourth_term}) we have expressed the
parametrization of Ref.~\cite{Belitsky:2005qn} in terms or our 4-vector and tensor conventions,
in such a way that the GPDs $G_{1,2,3,4}$ coincide with those of Ref.~\cite{Belitsky:2005qn}.

The BR parametrization uses four structures and so apparently solves the problem of the missing
structure of the GPV parametrization. However, it turns out that the fourth structure as defined in
Eq.~(\ref{eq:fourth_term}) is not independent but can be expressed as a linear combination of
the second and third,
\begin{align}
&\mathcal{K}^{\alpha \mu}_{3} n_{\mu}-\frac{\Delta^{2}}{2(\Delta \cdot P)}\mathcal{K}^{\alpha \mu}_{2} n_{\mu}
\nonumber \\
&= \mathcal{K}^{\alpha \mu }_{4} n_{\mu} \left[ 1 + \frac{\Delta^{2}}{2\xi (\Delta \cdot P)}\right].
\end{align}
A different choice is therefore needed to provide a complete parametrization of the transition matrix element.

\subsection{Complete parametrization}
\label{subsec:complete_vector}
Here we propose a complete parameterization of the $N \to \Delta$ transition matrix element
of the partonic QCD operator. We have performed a systematic analysis of all possible tensor structures
$\mathcal{K}^{\alpha\mu}_I$ in the matrix element of the partonic operator Eq.~(\ref{bilinear_partonic}),
taking into account the
discrete symmetries (parity, time-reversal, and hermiticity), the dynamical equations and constraints for
the $N$ and $\Delta$ spinors, and the relations resulting from the contraction of the tensors with
the light-cone vector $n^\mu$ (twist-2 projection). Based on this analysis we identify the
four independent structures
\begin{align}
n^{\alpha} \gamma^{\mu}  \gamma_{5}, \quad \Delta^{\alpha} \gamma^{\mu}\gamma_{5},
\quad  \Delta^{\alpha} P^{\mu}  \gamma_{5}, \quad g^{\alpha\mu}   \gamma_{5}.
\label{eq:possible_GPDs}
\end{align}
The first term was not included in previous studies. It is proportional to the light-cone vector $n$
and therefore does not appear in the decomposition of the matrix element of the local current operator,
not even when the requirement of current conservation is abandoned. However, it does appear in the
matrix element of the nonlocal partonic operator and corresponds to a GPD with vanishing first moment.

We define a new tensor structure as
\begin{align}
\mathcal{K}^{\alpha \mu}_{X} &\equiv m_{N} n^\alpha \gamma^{\mu} \gamma_{5}.
\label{Def_KX_traditional}
\end{align}
This structure can be used to complete the parametrization of
the transition matrix element in the form with multipole-type GPDs, Eq.~\eqref{eq:Goeke_para},
or in the form with covariant-type GPDs, Eq.~\eqref{eq:Belitsky_para} [excluding the
$\mathcal{K}^{\alpha \mu}_{4}$ structure of \eqref{eq:fourth_term}]. This provides a complete
GPD parametrizations with the traditional form of the tensor structures inherited from the
form factor parametrization, Eqs.~(\ref{eq:2}) and (\ref{eq:1}).

We can go further and construct a complete GPD parametrization with ``consistent'' tensor structures
satisfying the condition \eqref{eq:con}. Imposing this condition on the new structure
$n^{\alpha} \gamma^{\mu}  \gamma_{5}$ in Eq.~\eqref{eq:possible_GPDs}, we define the consistent
structure as 
\begin{align}
(\mathcal{K}^{\mathrm{con}}_X)^{\alpha \mu} &\equiv
m_{N} \left( n^{\alpha} - \frac{n \cdot p_{\Delta}}{p^{2}_{\Delta}} \, p^{\alpha}_{\Delta} \right)
\gamma^{\mu} \gamma_{5},
\label{Def_KX}
\end{align}
and use it to complement the consistent structures in the vector current matrix element,
Eq.~\eqref{eq:VdH_para} \cite{Pascalutsa:2006up,Semenov-Tian-Shansky:2023bsy}.
This leads to the following parametrization of the $N \to \Delta$ transition matrix element of the
vector partonic operator: 
\begin{align}
\mathcal{M}[\gamma]
&= C_{\mathrm{iso}} \sum_{I=M,E,C,X} \bar{u}_{\alpha}(\lambda_\Delta)
(\mathcal{K}^{\mathrm{con}}_{I})^{ \alpha \mu} n_{\mu} u(\lambda_N)
\nonumber \\[1ex]
&\times h_{I}(x,\xi,t)
\label{eq:new_para}
\end{align}
where the tensor structures are defined in Eqs.~\eqref{eq:VdH_para} and (\ref{Def_KX}),
and the GPDs satisfy the sum rules
\begin{subequations}
\begin{align}
\int^{1}_{-1}  dx \, h_{M,E,C}(x,\xi,t)&=  2g_{M,E,C}(t) ,
\\
\int^{1}_{-1}  dx \, h_{X}(x,\xi,t)&= 0.
\end{align}
\end{subequations}
Equation~\eqref{eq:new_para} provides a complete parametrization of the transition matrix element
of the partonic operator and definition of the transition GPDs based on consistent tensor structures.
\section{Axial vector transition GPDs}
\label{sec:axial}
\subsection{Adler et al.\ form factors}
We now extend the analysis of the $N\rightarrow \Delta$ transition matrix elements to the axial
vector partonic and local operators, which present similar issues as the vector operators.
The transition matrix element of the neutral local axial vector current operator,
with the same normalization as the vector current Eq.~\eqref{eq:mat_EM}, is defined as
\begin{align}
& \langle \Delta^+  | \bar{\psi}(0) \gamma^{\mu}  \gamma_5  \frac{\tau^{3}}{2} \psi(0) | p \rangle
= \frac{1}{2} \mathcal{J}^{\mu}[\gamma \gamma_{5}]_{p \rightarrow \Delta^+}.
\end{align}
The decomposition in bilinear forms is performed as in Eq.~(\ref{local_bilinear}).
As the parity counterpart of the vector current, the possible tensor structures are those of
Eq.~(\ref{eq:possible_EM}) with the $\gamma_{5}$ matrix dropped,
\begin{align}
 \Delta^{\alpha} \Delta^{\mu} , \quad \Delta^{\alpha} \gamma^{\mu}, \quad
 \Delta^{\alpha} P^{\mu} , \quad g^{\alpha\mu}.
\end{align}
Due to PCAC the isovector axial current is not conserved even in the limit of exact isospin symmetry.
Consequently, four independent axial-vector form factors appear in the $N \to \Delta$
transition. Following the parameterization by Adler et al.~\cite{Adler:1975tm}, we have
\begin{align}
&\langle \Delta^+  | \bar{\psi}(0) \gamma^{\mu}  \gamma_5  \frac{\tau^{3}}{2} \psi(0) | p \rangle
\nonumber \\
&= \sum_{I=5,6,3,4} \bar{u}_{\alpha}(\lambda_\Delta)  \tilde{\mathcal{K}}^{\alpha \mu}_{I} u(\lambda_N)
\, C^{A}_{I}(t),
\label{eq:axial_FFs}
\end{align}
where the Lorentz tensors are defined as
\begin{subequations}
\label{eq:axial_cov}
\begin{align}
\tilde{\mathcal{K}}^{\alpha \mu}_{5} &= g^{ \alpha \mu},
\\[1ex]
\tilde{\mathcal{K}}^{\alpha \mu}_{6} &= \frac{1}{m^{2}_{N}} \Delta^{ \alpha} \Delta^{ \mu},
\\
\tilde{\mathcal{K}}^{\alpha \mu}_{3} &=
\frac{1}{m_{N}}  [g^{ \alpha \mu} \slashed{\Delta} - \gamma^{\mu} \Delta^{\alpha}],
\\
\tilde{\mathcal{K}}^{\alpha \mu}_{4} &=
\frac{2}{m^{2}_{N}} [ g^{\alpha \mu} (P\cdot \Delta)  - P^{\mu} \Delta^{\alpha} ].
\end{align}
\end{subequations}
\subsection{GPV parametriztion}
A parametrization of the $N \rightarrow \Delta$ transition matrix element of the axial vector
partonic operator Eq.~(\ref{eq:mat_GPD}) was introduced by GPV \cite{Goeke:2001tz} based on
the Adler et al.\ form factors,
\begin{align}
\mathcal{M}[\gamma \gamma_5]
= C_{\mathrm{iso}} \sum_{I=5,6,3,4}
\bar{u}_{\alpha}(\lambda_\Delta) \tilde{\mathcal{K}}^{\alpha \mu}_{I} n_{\mu} u(\lambda_N) \,
C_{I}(x ,\xi , t) ,
\label{eq:axial_GPDs_para}
\end{align}
where the covariant structures $\tilde{\mathcal{K}}^{\alpha \mu}_{5,6,3,4}$ are those of
Eq.~\eqref{eq:axial_cov}. Note that $C_{5,6}(x,\xi,t)|_{\text{this work}}
= C_{1,2}(x,\xi,t)|_{\text{\cite{Goeke:2001tz, Semenov-Tian-Shansky:2023bsy}}}$.
The axial vector GPDs are related to the form factors by the sum rules
\begin{align}
\sqrt{\frac{2}{3}} \int^{1}_{-1}  dx \, C_{5,6,3,4}(x,\xi,t)= 2C^{A}_{5,6,3,4}(t).
\end{align}
However, similar to the situation with the vector operator, the tensor structures are not independent,
because
\begin{align}
\tilde{\mathcal{K}}^{\alpha \mu}_{4} n_{\mu}
= \frac{2}{ m^{2}_{N}} (P\cdot \Delta) \tilde{\mathcal{K}}^{\alpha \mu}_{5} n_{\mu}
+ \frac{1}{\xi} \tilde{\mathcal{K}}^{\alpha \mu}_{6} n_{\mu},
\label{eq:absorb}
\end{align}
where we have used the definition of $\xi$ in Eq.~(\ref{xi_t_def}). As a consequence, the
$C_4$ GPD can be absorbed in $C_5$ and $C_6$. This rearrangement does not introduce singularities
in $\xi$ because $\tilde{\mathcal{K}}^{\alpha \mu}_{6} n_{\mu}$ is proportional to $\xi$,
\begin{align}
\tilde{\mathcal{K}}^{\alpha \mu}_{6} n_{\mu} = - \frac{1}{m^{2}_{N}} \Delta^{\alpha} 2 \xi.
\end{align}
The parametrization Eq.~(\ref{eq:axial_GPDs_para}) thus contains only three independent
structures, one less than the number determined on general grounds, Eq.~\eqref{number_structures}.
An additional structure is therefore required.

\subsection{Complete parametrization}
\label{subsec:complete_axial}
To complete the parametrization of the transition matrix element of the axial vector partonic operator,
we proceed in the same way as in the vector case.
Based on a systematic analysis of the possible tensor structures,
taking into account the discrete symmetries, equations for the spinors, and the twist-2 projection, 
we identify the four independent structures
\begin{align}
n^{\alpha} \gamma^{\mu}  , \quad \Delta^{\alpha} \gamma^{\mu}, \quad  \Delta^{\alpha} P^{\mu}  ,
\quad g^{\alpha\mu}
\label{eq:possible_GPDs_axial}
\end{align}
As in the vector case, a structure proportional to the light-cone vector $n$ emerges and brings in a
GPD with vanishing first moment. We define the new tensor structure as
\begin{align}
\tilde{\mathcal{K}}^{\alpha \mu}_{X}
&=  m_{N} n^{\alpha}  \gamma^{\mu}.
\label{K_X_axial_cov}
\end{align}
In the parametrization Eq.~\eqref{eq:axial_GPDs_para} we then replace the redundant structure
$\mathcal{\tilde{K}}^{\alpha \mu}_{4}$ [see Eq.~\eqref{eq:absorb}] by the new independent
term ~\eqref{K_X_axial_cov}. This leads to the following parametrization of the $N \to \Delta$
transition matrix element of the axial vector partonic operator:
\begin{align}
\mathcal{M}[\gamma \gamma_5]
= C_{\mathrm{iso}} \sum_{I=5,6,3,X} \bar{u}_{\alpha}(\lambda_\Delta)
\tilde{\mathcal{K}}^{\alpha \mu}_{I} n_{\mu} u(\lambda_N) \, C_{I}(x,\xi,t) ,
\label{eq:com_axial_gpd}
\end{align}
where the tensor structures are defined in Eqs.~(\ref{eq:axial_cov}a,b,c) and \eqref{K_X_axial_cov}, 
and the GPDs satisfy the sum rules
\begin{subequations}
\begin{align}
&\sqrt{\frac{2}{3}} \int^{1}_{-1}  dx \, C_{5}(x,\xi,t)
\nonumber \\
&= 2\left[C^{A}_{5}(t) + \frac{m^{2}_{\Delta}-m^{2}_{N}}{m^{2}_{N}} C^{A}_{4}(t)\right], 
\\
&\sqrt{\frac{2}{3}} \int^{1}_{-1}  dx \, C_{6}(x,\xi,t)
= 2\left[C^{A}_{6}(t) + \frac{1}{\xi} C^{A}_{4}(t)\right],
\\
&\sqrt{\frac{2}{3}}\int^{1}_{-1}  dx \, C_{3}(x,\xi,t)
= 2C^{A}_{3}(t),
\\
&\sqrt{\frac{2}{3}}\int^{1}_{-1}  dx \, C_{X}(x,\xi,t)
= 0.
\end{align}
\end{subequations}
Alternatively, we can define a parametrization of the axial vector matrix element in terms of
consistent tensor structures, satisfying the condition Eq.~\eqref{eq:con} in analogy with 
the vector case. The new structure Eq.~(\ref{K_X_axial_cov}) can be extended to a consistent
structure as 
\begin{align}
(\tilde{\mathcal{K}}^{\mathrm{con}}_X)^{\alpha \mu}
&= m_{N} \left( n^{\alpha} - \frac{n \cdot p_{\Delta}}{p^{2}_{\Delta}} \, p^{\alpha}_{\Delta} \right)
\gamma^{\mu},
\label{K_X_axial}
\end{align}
and the other structures in Eq.~(\ref{eq:axial_cov}a,b,c) 
can be converted to consistent structures in a similar way.
\section{Interpretation and discussion}
\label{sec:interpretation}
\subsection{Light-front multipole expansion}
\label{subsec:multipole}
It is interesting to analyze the spin structure of the $N \rightarrow \Delta$ transition matrix element
of the partonic operators using light-front helicity states. This exercise confirms the presence of the
new structures observed in the 4D covariant decomposition and explains them in terms of
multipole transitions (dipole, quadrupole).

The analysis is performed using the standard light-front representation of the matrix element
\cite{Diehl:2003ny}. The light-like vector is taken along the 3-direction,
$n^\mu \propto (1, -\vec{e}_3)$, and
light-cone 4-vector components are defined as $v^{\pm} \equiv v^0 \pm v^3, \bm{v}_T \equiv (v^1, v^2)$.
The matrix element is considered in a frame where $\bm{P}_T = 0$ and $\bm{\Delta}_T \neq 0$,
so that the $N$ and $\Delta$ states have transverse momenta $-\bm{\Delta}_T/2$ and $\bm{\Delta}_T/2$,
respectively. The $N$ and $\Delta$ spin states are chosen as light-front helicity states \cite{Brodsky:1997de}.
The spinors are prepared in the particle rest frame, quantized along the $3$-direction, and boosted to the
desired longitudinal and transverse momentum by a sequence of longitudinal and transverse light-front boosts.
The spin states prepared in this way transform in simple way under light-front boosts
and exhibit a close analogy with nonrelativistic spin states.

In this representation we can discuss the spin structure of the matrix element Eq.~\eqref{eq:mat_GPD}
in terms of transitions between light-front helicity states
\begin{align}
\langle \textstyle{\frac{3}{2}} \lambda_\Delta | ... | \textstyle{\frac{1}{2}} \lambda_N \rangle .
\end{align}
The spin projections refer to a common quantization axis in the particle rest frames, and transitions
can be discussed in much the same way as for nonrelativistic systems. According to the rules of angular
momentum addition, a transition matrix element between spin-1/2 and spin-3/2 states is characterized
by a vector ($J = 1$) and tensor ($J = 2$) transition moment. Their spherical components are
given by the vector coupling coefficients
\begin{subequations}
\begin{align}
V^M &= \langle \textstyle{\frac{1}{2}} \lambda_N, 1 M|
\textstyle{\frac{3}{2}} \lambda_\Delta \rangle
& (M = 0, \pm 1),
\label{V_spherical}
\\
Q^M &= \langle \textstyle{\frac{1}{2}} \lambda_N , 2 M|
\textstyle{\frac{3}{2}} \lambda_\Delta \rangle
& (M = 0, \pm 1, \pm 2).
\label{Q_spherical}
\end{align}
\end{subequations}
Cartesian components are obtained using the standard rules \cite{Landau:1991wop},
\begin{subequations}
\begin{align}
&V^a & (a = 1,2,3),
\\
&Q^{ab} = Q^{ba}, \hspace{1em} Q^{aa} = 0 & (a,b = 1,2,3).
\end{align}
\end{subequations}
Because of the angular momentum difference of one unit between the 1/2 and 3/2 states,
$V$ is a true vector and $Q$ is a pseudotensor under parity. (For spin-1/2 to 1/2 transitions,
the transition vector is a pseudovector, and no transition tensor is allowed.)
The possible multipole structures in the matrix element Eq.~\eqref{eq:mat_GPD} arise as contractions
of the spin transition parameters $V$ and $Q$ with the longitudinal vector $\vec{e}_3$,
or the transverse momentum $\bm{\Delta}_T$. The structures are constrained by parity.
In the matrix element of vector partonic operator Eq.~\eqref{eq:mat_GPD},
the overall structures must be parity-odd. We obtain the following
multipole structures $(i,j = 1,2)$:
\begin{align}
\begin{array}{lrr}
\textrm{multipole} & M & \textrm{name} \\
\hline
V^i \epsilon^{3ij} \Delta_T^j & \pm 1 & V1 \\[1ex]
Q^{33} & 0 & Q0 \\[1ex]
Q^{3i} \Delta_T^i & \pm 1 & Q1 \\[1ex]
Q^{ij} (\Delta_T^i \Delta_T^j - \delta^{ij} |\bm{\Delta}_T|^2/2)
\hspace{2em} & \pm 2 & \hspace{2em}
Q2
\end{array}
\nonumber \\
\label{3D_multipoles}
\end{align}
where in the second column we indicate the light-front helicity difference
$M = \lambda_\Delta - \lambda_N$ in transitions mediated by the structure,
and in the third column name we give a name for later reference.
The total number of independent structures is four, in agreement with the number
established earlier using other methods (see Sec.~\ref{sec:transition_gpds}).

We can now compare the multipole structures in Eq.~(\ref{3D_multipoles}) with the bilinear
forms in the covariant decomposition of the matrix element in Eq.~(\ref{eq:possible_GPDs}).
Explicit expressions of the bispinor wave functions for the light-front helicity states are
given in Refs.~\cite{Lorce:2017isp,Granados:2016jjl}. Evaluating the bilinear forms with the
4-vector components of $P, \Delta$ and $n$ as specified above, inspecting the allowed light-front
helicity transitions, and comparing them with the transitions allowed by the
multipoles Eq.~(\ref{3D_multipoles}), we obtain the matrix
\begin{align}
\begin{array}{r|cccc}
 & V1 & Q0 & Q1 & Q2 \\[1ex]
 \hline
\Delta^{\alpha} \gamma^{\mu}\gamma_{5} &\checkmark &\checkmark &\checkmark & \\[1ex]
\Delta^{\alpha} P^{\mu}  \gamma_{5} &\checkmark &\checkmark &\checkmark &\checkmark \\[1ex]
g^{\alpha\mu}   \gamma_{5} &\checkmark &\checkmark &\checkmark & \\[1ex]
n^{\alpha} \gamma^{\mu} \gamma_{5} & &\checkmark & & \\[1ex]
\end{array}
\end{align}
One observes that new covariant structure $n^{\alpha} \gamma^{\mu} \gamma_{5}$
projects on the $M = 0$ spin quadrupole transition $Q0$.
While the other bilinears also project on this multipole, the fourth structure is needed so that all
multipoles can be obtained as combinations of the bilinear forms. When the fourth structure is included,
the matrix connecting the covariant structures and the multipoles is nonsingular,
and the relation can be inverted.

The analysis can be extended to the matrix element of the axial vector partonic operator.
The spin transition vector and pseudotensor, Eqs.~(\ref{V_spherical}) and (\ref{Q_spherical}),
are the same as in the case of the vector partonic operator, but the matrix element must now be
overall parity-even. We obtain the following structures:
\begin{align}
\begin{array}{lrr}
\textrm{multipole} & M & \textrm{name} \\
\hline
V^3 & 0 & V0 \\[1ex]
V^i \Delta_T^i & \pm 1 & V1 \\[1ex]
Q^{3i} \epsilon^{3ij} \Delta_T^j & \pm 1 & Q1 \\[1ex]
Q^{ij} \epsilon^{3jk} \Delta_T^i \Delta_T^k \hspace{1em} & \pm 2  & \hspace{2em}
Q2
\end{array}
\nonumber \\
\label{3D_multipoles_axial}
\end{align}
Note that there are now two vector and two quadrupole structures. 
The total number of structures is again four, in agreement with the number established earlier.
Comparing the multipole structures with the covariant bilinear forms, we now obtain the matrix
\begin{align}
\begin{array}{r|cccc}
& V0 & V1 & Q1 & Q2 \\[1ex]
\hline
\Delta^{\alpha} \gamma^{\mu} &\checkmark &\checkmark & & \\[1ex]
\Delta^{\alpha} P^{\mu}  &\checkmark &\checkmark &\checkmark &\checkmark \\[1ex]
g^{\alpha\mu}  &\checkmark & \checkmark &\checkmark & \\[1ex]
n^{\alpha} \gamma^{\mu} & \checkmark & & & \\[1ex]
\end{array}
\end{align}
One observes that new covariant structure $n^{\alpha} \gamma^{\mu}$ projects on the
$M = 0$ spin dipole transition $V0$. Again the new structure is needed so that all multipoles
can be obtained as combinations of the bilinear forms.
\subsection{$1/N_c$ expansion}
\label{subsec:nc_expansion}
The $1/N_c$ expansion is a powerful method for analyzing $N \rightarrow \Delta$ transition matrix elements
of QCD operators and has been applied extensively to transition form factors and GPDs \cite{Goeke:2001tz}.
We want to comment briefly on the additional structure in the parametrization of the matrix element
from the perspective of the $1/N_c$ expansion.

The leading structures in the $1/N_c$ expansion of the $N \rightarrow \Delta$ transition GPDs are the
spin dipole transitions involving the spin vector ${V}$, Eq.~(\ref{V_spherical}). These transitions
are favored by ``$I = J$ rule'' connecting the $t$-channel isospin and spin, 
which follows from the contracted spin-flavor symmetry of baryons in the large-$N_c$ limit.
The spin quadrupole transitions involving
the spin tensor $Q$, Eq.~(\ref{Q_spherical}), are subleading in the $1/N_c$ expansion.

In the vector transition matrix element, the new structure introduced in Sec.~\ref{subsec:complete_vector}
is a spin quadrupole [see Eq.~\eqref{3D_multipoles}] and therefore subleading in the $1/N_c$ expansion.
As such its contribution is likely numerically small. In this sense the omission of this structure
in the original GPV parametrization may be justified on numerical grounds.
Notice that the GPD associated with the new structure cannot be inferred by connecting
$N \rightarrow \Delta$ and $N \rightarrow N$ matrix elements through the spin-flavor symmetry,
as the quadrupole structure exists only in $N \rightarrow \Delta$ transitions and has
no correspondence in $N \rightarrow N$. The GPDs associated with the new structure can
therefore only be estimated using dynamical models.

In the axial vector transition matrix element, the new structure introduced in Sec.~\ref{subsec:complete_axial}
is a spin dipole transition and thus appears in leading order of the $1/N_c$ expansion.
As such, this structure could also be numerically large. This needs to be studied
in a careful analysis, separating the contributions of the $V0$ structures from the other multipoles
in the large-$N_c$ limit.
\subsection{Cross-channel partial wave analysis}
\label{subsec:cross_channel}
The need for the additional covariant structure in the decomposition of the
$N \rightarrow \Delta$ matrix element can also be demonstrated independently
from the perspective of the cross-channel SO$(3)$ partial-wave (PW) analysis.
This analysis relies on the fact that the number of independent structures characterizing
an amplitude is the same in all channels related by crossing \cite{Berestetskii:1982qgu}.

We consider the matrix element of the partonic operator for transitions between the vacuum
and a $\Delta \bar N$ (antinucleon) state,
\begin{align}
\langle \Delta \bar N | \bar{\psi}(-\tau n/2) \Gamma^\mu n_\mu \hat{T} \psi(\tau n/2) | 0 \rangle ,
\label{eq:me_cross}
\end{align}
where $\Delta$ and $\bar N$ have 4-momenta $p_\Delta$ and $p_{\bar N}$ and spin quantum numbers
$\lambda_\Delta$ and $\lambda_{\bar N}$. This matrix element is related to the $N \rightarrow \Delta$
transition matrix element Eq.~(\ref{eq:mat_GPD}) by crossing. The operation
involves regarding the matrix element as a function of the 4-momenta
$p_N$ and $p_\Delta$, substituting $p_N = -  p_{\bar N}$,
which exchanges role of $P$ and $\Delta$ in Eq.~(\ref{P_Delta}),
\begin{align}
  \Delta = p_\Delta + p_{\bar N}, \hspace{2em}
  P = (p_\Delta - p_{\bar N})/2,
\label{Delta_P_cross}
\end{align}
and analytically continuing the dependence on the invariant variables. Here we consider
Eq.~(\ref{eq:me_cross}) in the physical region of the $t$-channel process,
\begin{align}
t = \Delta^2  
\ge (m_\Delta + m_N)^2.
\label{tchannel_physical}
\end{align}
In this region it describes the creation of a $\Delta \bar N$ pair by the QCD operator, as would
happen e.g. in the exclusive process $\gamma^\ast + \gamma \rightarrow \Delta + \bar N$.
Partonic variables appropriate for the $t$-channel physical region can be defined, and the
matrix element can be described in terms of so-called generalized distribution amplitudes (GDAs)
(see e.g.\ Refs.\cite{Diehl:1998dk,Polyakov:1998ze} for the simplest case of $\pi \pi$ GDAs);
in the present study the focus is on the spin structure of the matrix element Eq.~(\ref{eq:me_cross}),
and the $t$-channel partonic variables will not be needed explicitly. 

The spin structure of the matrix element Eq.~(\ref{eq:me_cross}) can be exhibited through a
partial-wave expansion in the $t$-channel. At the same time, the matrix element can be expanded
in covariant structures in the same way as the $s$-channel matrix element Eq.~(\ref{eq:mat_GPD}).
The covariant structures in both channels are related by crossing. This will allow us to match
the $s$-channel covariant structures with the $t$-channel partial waves and validate the
number of independent structures.

To perform the partial-wave expansion, we use the center-of-mass frame of the $\Delta \bar N$ system
and choose a coordinate system such that the light-like vector is along 3-direction $\vec{n} \parallel \vec{e}_3$.
The orbital motion of the $\Delta \bar N$ system is described by the center-of-mass momentum
$\vec{p} \equiv \vec{p}_\Delta = -\vec{p}_{\bar N}$, and the angle relative to the 3-axis is defined as
\begin{align}
\cos\theta_t \equiv \vec{p}\cdot\vec{n} / |\vec{p}|,
\hspace{2em} -1 \leq \cos\theta_t \leq 1.
\end{align}
The $\Delta$ and $\bar N$ spin degrees of freedom described by the helicities $\lambda_\Delta$
and $\lambda_{\bar N}$, defined as the spin projections on the momenta $\vec{p}$ and $-\vec{p}$
(here we are using canonical helicity, not light-front helicity). The orbital motion of the
$\Delta\bar N$ system can then be expanded in helicity partial waves with total angular momentum
$J$ and projection $J_3$, whose wave functions given by \cite{Berestetskii:1982qgu}
\begin{align}
&\Phi_{J J_3, \lambda_\Delta \lambda_{\bar N}} (\vec{p}) = \sqrt{\frac{2J + 1}{4\pi}}
D^{(J)}_{\Lambda J_3}(\vec{p}/|\vec{p}|) S_{\lambda_\Delta \lambda_{\bar N}},
\nonumber \\
&\Lambda \equiv \lambda_\Delta - \lambda_{\bar N}, \hspace{2em} J = 0,1,2...
\end{align}
where $D^{(J)}_{\Lambda J_3}$ are the finite-rotation matrices and $S_{\lambda_\Delta \lambda_{\bar N}}$
denotes the spin wave function of the $\Delta\bar N$
system in the body-fixed frame aligned with $\vec{p}$ axis.

Following the analysis presented in Sec.~4.2 of Ref.~\cite{Diehl:2003ny}, we
now consider the matrix elements of the local twist-2 operators
obtained by expanding partonic operator Eq.~(\ref{eq:me_cross})
in powers of the light-like distance $\tau$ (here $k = 0, 1, 2...$)
\begin{align}
\langle \Delta \bar N | \, \bm{S} \, \bar{\psi}(0) \Gamma^\mu
\overset{\leftrightarrow}{D} \vphantom{D}^{\mu_1}\cdots \overset{\leftrightarrow}{D}\vphantom{D}^{\mu_k}
\hat T \psi(0) \, | 0 \rangle \; n_{\mu} n_{\mu_1} \ldots n_{\mu_k},
\label{Matrix_el}
\end{align}
where $\bm{S}$ denotes the symmetrization of the 4-tensor indices and subtraction of the traces.
We decompose the matrix element Eq.~(\ref{Matrix_el}) it in the $t$-channel partial waves
with total angular momentum $J$. Because all tensor indices in Eq.~(\ref{Matrix_el}) are contracted
with the vector $n$ and thus restricted to values $0$ or $3$, the PWs with total angular momentum
$J$ all have $J_3=0$, and their angular dependence is described by the rotation functions
$d^{(J)}_{J_3 |\Lambda|}(\theta_t)$, where $\Lambda = \lambda_\Delta- \lambda_{\bar N}$
(see {\it e.g.} Appendix A.2 of Ref.~\cite{Martin:102663}).
As a result we obtain the following three structures in the angular dependence,
depending on the coupling of the $\Delta$ and $\bar N$ helicities:
\begin{align}
\begin{array}{cl}
|\lambda_\Delta - \lambda_{\bar N}| \hspace{1em} & \textrm{angular dependence}
\\[1ex]
0 & d^{(J)}_{00}(\theta_t) \propto P_J(\cos \theta_t) \\[1ex]
1 & d^{(J)}_{01}(\theta_t) \propto \sin \theta_t P'_J(\cos \theta_t) \\[1ex]
2 & d^{(J)}_{02}(\theta_t) \propto \sin^2 \theta_t P^{''}_J(\cos \theta_t)
\end{array}
\label{angular_dependences}
\end{align}
where $P_{J}$ are the Legendre polynomials and the prime denotes the derivative with respect to the argument.
Note in particular that the $|\lambda_\Delta - \lambda_{\bar N}|=2$ helicity state gives rise to an angular
dependence proportional to the second derivatives of the Legendre polynomials of $\cos \theta_t$.

We can now confront the angular dependence obtained from the $t$-channel partial-wave expansion
with that contained in the covariant decomposition of the matrix element.
The covariant decomposition of the $t$-channel matrix element Eq.~(\ref{eq:me_cross})
is obtained by performing the crossing operation in the bilinear forms of the $s$-channel
matrix element of Sec.~\ref{subsec:complete_vector}, using Eq.~(\ref{Delta_P_cross})
for the $t$-channel 4-vectors (the baryon spinors and the covariant structures depend
polynomially on the 4-momenta, and no analytic continuation is involved here).
Comparing the angular dependence, we find that the
$P''_J(\cos \theta_t)$ dependence of the $|\lambda_\Delta- \lambda_{\bar N}|=2$ term in
Eq.~(\ref{angular_dependences}) requires the presence of all four covariant structures
in Eq.~(\ref{eq:possible_GPDs}), including the new structure $n^{\alpha} \gamma^{\mu} \gamma_{5}$,
\begin{align}
P''_J(\cos \theta_t) \; \leftrightarrow \;  \{ {\cal K}_X, {\cal K}_M, {\cal K}_E, {\cal K}_C\}.
\end{align}
The structures ${\cal K}_{M,E,C}$ alone, Eq.~(\ref{eq:VdH_para}), or amended by
the structure ${\cal K}_{4}$, Eq.~(\ref{eq:fourth_term}), result in
a parametrization of the cross-channel matrix element that is incomplete from the
perspective of the $t$-channel partial-wave analysis. This confirms the necessity of the
four independent structures, Eq.~(\ref{eq:possible_GPDs}), obtained in the direct analysis
of the $s$-channel matrix element.

The same $t$-channel partial wave analysis can be applied to the matrix element of the axial vector
partonic operator and confirms the necessity of including the tensor structure
$\tilde{{\cal K}}_{X}$, Eq.~(\ref{K_X_axial_cov}) in the covariant decomposition.

In the present study we use the $t$-channel partial wave expansion only for validating the covariant
structures in in the decomposition of the matrix element, remaining in the
physical region of the $t$-channel process, Eq.~(\ref{tchannel_physical}).
The $t$-channel partial wave expansion
can also be analytically continued to the $s$-channel physical region to construct a partial wave
representation of the GPDs. For $N \rightarrow N$ GPDs this has been realized
in the framework of the dual parametrization of GPDs \cite{Polyakov:2002wz,Muller:2014wxa},
an approach based on the Mellin-Barnes integral technique \cite{Kumericki:2007sa},
and in the so-called universal moment parametrization (GUMP) \cite{Guo:2022upw,Guo:2023ahv}.
These approaches combine the SO(3) partial wave expansion with the conformal moment expansion
of the GPDs, which diagonalizes the QCD evolution equations and enables an effective
implementation of the scale dependence. Extending these formulations to $N \rightarrow \Delta$
transition GPDs would be an interesting further development.

\section{Conclusions and extensions}
\label{sec:summary}
In this work we have studied the structural decomposition of the $N \rightarrow \Delta$ transition
matrix elements of the partonic QCD operators and the corresponding definition of the transition GPDs.
The results can be summarized as follows:

\begin{itemize}

\item[(i)]
The covariant decomposition of the $N \rightarrow\Delta$ matrix element of the vector partonic operator
requires a new structure in addition to the three structures present in the matrix element of the
local vector current operator, see Eq.~(\ref{eq:possible_GPDs}).
The new structure is proportional to the light-like 4-vector
and specific to the partonic operator. It corresponds to a transition GPD with zero
first moment.

\item[(ii)]
With the new structure included, a complete parametrization of the $N\rightarrow \Delta$
matrix element of the vector partonic operator in terms of four transition GPDs is achieved.
The previous parametrizations in the literature \cite{Goeke:2001tz,Belitsky:2005qn}
use too few or linearly dependent structures and are incomplete.

\item[(iii)]
The presence of the new structure in the covariant decomposition is confirmed by the light-front
multipole expansion of the matrix element. The new structure describes $M = 0$ spin quadrupole
transitions enabled by the transverse momentum transfer. The need for the new structure is also
demonstrated by the $t$-channel partial wave expansion of the matrix element.

\item[(iv)]
A similar new structure and GPD with zero moment appear in the $N \rightarrow \Delta$ transition
matrix element of the axial vector partonic operator, see Eq.~(\ref{eq:possible_GPDs_axial}).
With this structure included, a complete
parametrization of the axial vector matrix element in terms of four transition GPDs is achieved.
The new structure in the axial vector matrix element describes $M = 0$ spin dipole transitions.

\end{itemize}
These results provide a basis for calculations of the $N \rightarrow \Delta$ transition GPDs
from nonperturbative dynamics (lattice QCD, chiral effective field theory, composite models)
and the analysis of hard exclusive processes with $N \rightarrow \Delta$ transitions
(deeply-virtual Compton scattering, chiral-even meson production processes).

The studies reported here could be extended in several directions.
The new structure in the $N \rightarrow \Delta$ transition matrix element of the partonic operator
identified in the present study is accompanied by a transition GPD of vanishing first moment.
However, its second moment can be non-zero. As such the new structure can influence the
$N \rightarrow \Delta$ transition matrix elements of the quark flavor components of the QCD
energy-momentum tensor \cite{Kim:2022bwn, Kim:2023xvw}. Its effect in this context should be investigated.

The $1/N_c$ expansion is the primary tool for analyzing and modeling the $N \rightarrow \Delta$
transition GPDs. Our arguments in Sec.~\ref{subsec:nc_expansion} suggest that the new structure
in the $N \rightarrow \Delta$ matrix element of the vector partonic operator is subleading in $1/N_c$,
while in the axial vector operator it is leading in $1/N_c$. These findings should be validated
by a comprehensive large-$N_c$ analysis of the transition GPDs. The $1/N_c$ expansion can also
suggest natural definitions of the independent transition GPDs, as multipole structures subject
to the $I = J$ rule \cite{Kim:2024ibz}.

Numerical studies of the $N \rightarrow \Delta$ transition GPDs can be performed with the
chiral quark-soliton model, which realizes the mean-field picture of baryons in the large $N_c$
limit with the effective dynamics arising from the spontaneous breaking of chiral symmetry.
Estimating the size of the new structure in the vector transition GPDs requires calculation
of structures at subleading order of the $1/N_c$ expansion, which appear due to collective
rotations of the mean-field system. Calculations are in progress \cite{inprep}.

In addition to the chiral-even partonic operators (vector, axial vector) considered in the present
study, the chiral-odd partonic operators (tensor) are needed to describe exclusive pion production
and other chirality-flipping processes with $N \rightarrow \Delta$ transitions \cite{Kroll:2022roq}.
A structural analysis of the chiral-odd $N \rightarrow \Delta$ GPDs should be performed
along the same lines as for the chiral-even ones here.

\section*{Acknowledgments}
This material is based upon work supported by the U.S.~Department of Energy, Office of Science,
Office of Nuclear Physics under contract DE-AC05-06OR23177. 

The research reported here takes place in the context of the Topical Collaboration ``3D quark-gluon
structure of hadrons: mass, spin, tomography'' (Quark-Gluon Tomography Collaboration) supported by
the U.S.~Department of Energy, Office of Science, Office of Nuclear Physics under
contract DE-SC0023646.

The research reported here is supported by the Basic Science Research Program through the National
Research Foundation of Korea (NRF) funded by the Ministry of Education RS-2023-00238703;
and under Grant No. NRF-2018R1A6A1A06024970 (Basic Science Research Program) (K.S., S.S.).
It is also supported by the France Excellence scholarship through Campus France funded by the French
government (Minist\`ere de l’Europe et des Affaires \'etrang\`eres), Grant-No.\ 141295X (H-Y.W.).

\appendix
\section{Conventions}
\label{app:conventions}
In this appendix we explain the correspondence between the bilinear forms in the
$N \rightarrow \Delta$ transition matrix elements obtained with different conventions
for the antisymmetric tensor.
Our work and Ref.~\cite{Belitsky:2005qn} use the convention of Eq.~(\ref{eq:conv_1})
for the antisymmetric tensor. References~\cite{Pascalutsa:2006up, Semenov-Tian-Shansky:2023bsy,
Goeke:2001tz} use instead the convention
\begin{align}
\epsilon_{0123}=-\epsilon^{0123} = 1,
\label{eq:conv_2}
\end{align}
which differs from Eq.~(\ref{eq:conv_1}) by a minus sign. With the convention Eq.~(\ref{eq:conv_2}),
the tensor relations corresponding to Eq.~\eqref{eq:Sign} take the form
\begin{subequations}
\label{eq:Sign_1}
\begin{align}
&i \epsilon^{\alpha  \mu P \Delta} \doteq m_{\Delta} m_{N} \mathcal{K}^{\alpha  \mu}_{1}
- \frac{m^{2}_{N}}{2}\mathcal{K}^{\alpha  \mu}_{2}-\frac{m^{2}_{N}}{2} \mathcal{K}^{\alpha  \mu}_{3},
\label{eq:Sign_1_a}
\\
&\epsilon^{\alpha \ P \Delta}_{\ \lambda}\epsilon^{\mu \lambda P \Delta} \gamma_{5} \cr
&\doteq -(\Delta \cdot p_{\Delta}) \frac{m^{2}_{N}}{2} \mathcal{K}^{\alpha \mu}_{2}
+ (P \cdot p_\Delta) m^{2}_{N} \mathcal{K}^{\alpha \mu}_{3},
\label{eq:Sign_1_b}
\\
&\Delta^{\alpha }(\Delta^{2} P^{\mu} - (\Delta \cdot P) \Delta^{\mu}) \gamma_{5}
\nonumber \\
& \doteq \Delta^{2} \frac{m^{2}_{N}}{2} \mathcal{K}^{\alpha \mu}_{2}
- (P \cdot \Delta) m^{2}_{N} \mathcal{K}^{\alpha \mu}_{3}.
\label{eq:Sign_1_c}
\end{align}
\end{subequations}
Equation~(\ref{eq:Sign_1}a) differs from Eq.~(\ref{eq:Sign}a) by a minus sign,
while Eqs.~(\ref{eq:Sign_1}b,c) remain the same as Eqs.~(\ref{eq:Sign}b,c).
Note that these relations hold true even when lowering the indices,
$i \epsilon^{\alpha \mu P \Delta} \to i \epsilon_{\alpha \mu P \Delta}$ etc.
Based on these relations, we obtain the following relations between the bispinor
matrices of Eq.~(\ref{eq:1}) in the two conventions:
\begin{align}
{\mathcal{K}}_{M}^{\alpha \mu} |_{\text{convention~\eqref{eq:conv_1}}}
&= -{\mathcal{K}}_{M}^{\alpha \mu} |_{\text{convention~\eqref{eq:conv_2}}},
\nonumber \\
{\mathcal{K}}_{E,C}^{\alpha \mu} |_{\text{convention~\eqref{eq:conv_1}}}
&= {\mathcal{K}}_{E,C}^{\alpha \mu} |_{\text{convention~\eqref{eq:conv_2}}}.
\label{eq:1_app}
\end{align}
The same relations apply to the consistent tensor structures of Eq.~(\ref{eq:VdH_para}):
\begin{align}
(\mathcal{K}^{\mathrm{con}}_M)^{\alpha \mu} |_{\text{convention~\eqref{eq:conv_1}}}
&= -(\mathcal{K}^{\mathrm{con}}_{M})^{\alpha \mu} |_{\text{convention~\eqref{eq:conv_2}}},
\nonumber \\
(\mathcal{K}^{\mathrm{con}}_{E,C})^{\alpha \mu}|_{\text{convention~\eqref{eq:conv_1}}}
&= (\mathcal{K}^{\mathrm{con}}_{E,C})^{\alpha \mu} |_{\text{convention~\eqref{eq:conv_2}}}.
\label{eq:2_app}
\end{align}
The form factors $G^{*}_{M,E,C}$ and $g_{M,E,C}$ remain the same in the two conventions
if the tensors in the bilinear forms are exchanged according to Eqs.~(\ref{eq:1_app}) and (\ref{eq:2_app}).
The conversion matrices connecting the different types of form factors (covariant, multipole),
Eqs.~\eqref{eq:connect}, \eqref{eq:connect_2}, and \eqref{eq:connect_3}, also remain the same.

Reference~\cite{Pascalutsa:2006up} uses the convention Eq.~\eqref{eq:conv_2} and defines
the decomposition of the matrix element with an overall minus sign compared to
Eq.~(\ref{eq:Jones_ME}),
\begin{align}
&\langle \Delta^{+} |J^{\mu}_{\mathrm{em}}(0) | p  \rangle
\nonumber \\[1ex]
&= \sqrt{\frac{2}{3}}
\sum_{I=M,E,C} \bar{u}_{\alpha} (\lambda_{\Delta})  
(-\mathcal{K}_{I}^{ \alpha \mu} |_{\text{\cite{Pascalutsa:2006up}}})   u(\lambda_{N}) \, G^{*}_{I}(t),
\end{align}
but defines the bispinor matrices with an overall minus compared to Eq.~(\ref{eq:1})
[evaluated with the convention Eq.~\eqref{eq:conv_2}]
\begin{align}
\mathcal{K}_{M,E,C}^{\alpha \mu} |_{\text{convention~\eqref{eq:conv_2}}}
&= -{ \mathcal{K}}_{M,E,C}^{\alpha \mu} |_{\text{\cite{Pascalutsa:2006up}}},
\end{align}
so that the two minus signs compensate each other. The form factors $G^{*}_{M,E,C}$ and
$g_{M,E,C}$ are the same in Ref.~\cite{Pascalutsa:2006up} and in our work.
\bibliography{ndelta_parametrization}
\end{document}